\documentclass[twocolumn,groupedaddress,showpacs,preprintnumbers,amsmath,amssymb]{revtex4}

\usepackage{graphicx}
\usepackage{dcolumn}
\usepackage{bm}

\usepackage{amsmath}
\usepackage{amssymb}
\usepackage{amsfonts}
\usepackage{mathrsfs}
\usepackage{epsfig}
\usepackage{bm}

\begin{document}

\title{Transverse self-modulation of ultra-relativistic lepton beams in the plasma wakefield accelerator}

\author{J. Vieira$^1$, Y. Fang$^2$, W.B. Mori$^3$, L. O. Silva$^1$ and P. Muggli$^{2,4}$}
\affiliation{$^1$GoLP/Instituto de Plasmas e Fus\~ao Nuclear - Laborat\'orio Associado Instituto Superior T\'ecnico (IST), Technical University of Lisbon, Lisboa, Portugal\\$^2$University of Southern California, Los Angeles, CA 90089, USA\\$^3$University of California, Los Angeles, CA 90095, USA\\ $^4$Max Planck Institute for Physics, Munich, Germany}

\begin{abstract}
The transverse self-modulation of ultra-relativistic, long lepton bunches in high-density plasmas is explored through full-scale particle-in-cell simulations. We demonstrate that long SLAC-type electron and positron bunches can become strongly radially self-modulated over centimeter distances, leading to wake excitation in the blowout regime with accelerating fields in excess of 20 GV/m. We show that particles energy variations exceeding 10 GeV can occur in meter-long plasmas. We find that the self-modulation of positively and negatively charged bunches differ when the blowout is reached. Seeding the self-modulation instability mitigates the effect of the competing hosing instability. This work reveals that a proof-of-principle experiment to test the physics of bunch self-modulation can be performed with available lepton bunches and with existing experimental apparatus and diagnostics.
\end{abstract}


\maketitle

\section{\label{sec:intro}Introduction}

Plasma-based acceleration offers the possibility to accelerate electrons and positrons in linear accelerators to high energies in distances much shorter than those achievable with conventional radio-frequency accelerating devices. Laser pulse (laser wakefield acceleration or LWFA~\cite{bib:tajima_prl_1979}) or particle bunch (plasma wakefield acceleration or PWFA~\cite{bib:chen_prl_1985}) drivers can excite large amplitude plasma waves or wakes. Peak electric fields in excess of $E_0 = 96~\sqrt{n_0 [\mathrm{cm}^{-3}]}~\mathrm{[V/\mathrm{m}]}$ can be generated, and accelerating gradients in excess of 100 GV/m can be achieved using plasma densities $n_0$ in the $10^{18}~\mathrm{cm^{-3}}$ range, typical of current experiments. Plasmas may thus contribute to a novel generation of compact accelerators, with impact on many scientific and technological applications~\cite{bib:patel_nature_2007}.

Proton bunches are attractive for plasma wakefield acceleration because the total amount of energy they carry (more than 100 kJ for a single LHC bunch) and the energy per particle (7 TeV at LHC) is much larger than that of the lepton bunches of a future linear collider ($1.6~\mathrm{kJ}$ for a typical $2\times10^{10}$, $500~\mathrm{GeV}$ lepton bunch). One can therefore envisage accelerating a single $\approx10~\mathrm{GeV}$ incoming lepton bunch to the TeV energy scale in a single PWFA stage driven by a relativistic proton bunch. The initial proposal for a proton-driven PWFA (or PDPWFA) \cite{bib:caldwell_natphys_2009} considered the case of a short ($\approx100\mu$m) proton bunch driver. However, such short proton bunches are not currently available. Kumar \emph{et al.}~\cite{bib:kumar_prl_2010} recently suggested that the radial self-modulation (S-M) of a long proton bunch at the wavelength of the relativistic plasma wave can lead to the resonant excitation of large amplitude plasma wakefields. This is analogous to the self-modulated LWFA~\cite{bib:esarey_prl_1996} that was used in early experiments when fs-long laser pulses were not available. The self-modulation of a particle bunch is completely analogous to that of a laser pulse (photon beam)~\cite{bib:mori_ieee_1997}. Studying experimentally the physics of the S-M of particle bunches that could be considered as drivers for large energy gain PWFA experiments is therefore interesting and important.

The S-M is the result of a transverse two-stream instability. For this instability to develop for high energy-beams the bunch transverse dimensions must be on the order of, or smaller than the cold plasma collisionless skin depth ($\sigma_r\approx c/\omega_p$) to avoid possible transverse bunch filamentation \cite{bib:fil}. Since the S-M is a convective instability, longer bunches, which encompass a higher number of plasma skin depths ($\sigma_{\xi}\gg c/\omega_p$) and longer propagation distances ($z\gg k_{\beta}^{-1}$) lead to larger wakefield amplitudes. Here $c$ is the speed of light, $\omega_p = \sqrt{n_0 e^2/(m_e \epsilon_0)}$ the electron plasma frequency, $m_e$ the electron mass, $e$ the elementary charge, $\epsilon_0$ the vacuum dielectric constant, $z$ the propagation distance, $k_{\beta}=\omega_p/(c \sqrt{2}\gamma_0)(m_e/m_p)^{1/2}$ the betatron wavenumber with $m_p$ the bunch particles' mass, $\gamma_0 m_p c^2$ the energy of each beam particle, $\gamma_0$ the relativistic factor, and $\sigma_r$ and $\sigma_{\xi}$ the bunch rms radius and length, respectively.

In the simulations presented here, we consider the case of high-energy electron and positron bunches that are currently available for experiments. We find that with these ultra-relativistic ($\gamma_0\propto40,000$) and long ($\sigma_{\xi}\propto500~\mu \mathrm{m}$) lepton bunches available at SLAC FACET \cite{bib:hogan_njp_2010} the saturation of the S-M instability is reached over only a few centimeters of plasma at a density of $2.3\times 10^{17}~\mathrm{cm}^{-3}$. This is due to the lower relativistic mass (25 GeV) of leptons when compared to that of protons in available bunches (450 GeV, See Table~\ref{tab:parameters}). These lepton beams can therefore be focused tightly and large plasma densities can be used ($\sigma_r\approx c/\omega_p\propto n_e^{-1/2}$), which lead to large transverse focusing and longitudinal accelerating wakefield amplitudes. In addition the betatron wavelength ($2\pi/k_\beta$) is much shorter so shorter propagation distances are needed.

In 3D numerical simulations we find that, as was previously noted~\cite{bib:whittum_prl_1991}, hosing-like instability of the bunch that is long when compared to the plasma wavelength may be a serious limitation to its stable propagation over distances longer than the saturation length of the S-M instability. The hosing also occurs at a longer wavelength than that of the wake. Similar behavior was observed in simulations of long laser pulses~\cite{bib:duda_prl_1999,bib:duda_pre_2000}. In an accelerator scheme this may also limit the energy gain by a short, externally injected witness bunch. However, these simulations also indicate that the seeding of the S-M instability can significantly reduce the growth of the hosing instability over the plasma lengths considered here, as was noted in the laser pulse case~\cite{bib:duda_prl_1999,bib:duda_pre_2000}. We consider the case of an electron bunch with an asymmetric current profile with rapidly rising edge to seed the S-M instability and defer to another publication the detailed study of the competition between the two instabilities. Further simulations with the shaped bunch in 2D cylindrically symmetric geometry that precludes hosing-like instabilities were performed to scan parameters. At saturation of the instability, the blowout regime \cite{bib:lu_2006} is reached and accelerating fields in excess of $20~$GV/m are excited. However, in this nonlinear regime the behavior of electron and positron bunches differ, in contrast with the linear regime for which their behavior is identical. Differences in wakefield excitation by short, positively and negatively charged bunches were described previously~\cite{bib:lu_pop_2005}. Because most of the volume of plasma accelerating structure in the nonlinear regime (ion column) is defocusing for positrons, a majority of positrons are quickly and strongly defocused and do not participate in the wakefields excitation. With the stable propagation of the electron and positron bunches over a meter-scale plasma, peak energy gain and loss by the bunch particles in the multi-GeV range are observed.

This paper is organized as follows. In Sec.~\ref{sec:lep_vs_had} we point out the differences between proton and lepton bunch S-M, and describe a possible S-M experiment using lepton bunches currently available at SLAC-FACET. In Sec.~\ref{sec:sim_param} we describe the base line parameters for the 3D and for the 2D cylindrically symmetric simulations. In Sec.~\ref{sec:hosing_vs_sm} we present 3D simulation results illustrating the competition between the hosing-like and the S-M instabilities with symmetric (with long rise time) and asymmetric (with rapid rise time) current profiles. In Sec.~\ref{sec:ep} we show 2D cylindrically symmetric simulation results of of the self-modulation of long SLAC electron and positron bunches. In Sec.~\ref{sec:wake}, we describe the S-M wakefields excitation over meter long plasmas. In Sec.~\ref{sec:challenges} we refer to possible diagnostics to observe the S-M. Finally we present some conclusions in Sec.~\ref{sec:conclusions}.

\section{Self-modulation of lepton and proton bunches}
\label{sec:lep_vs_had}

The growth of the S-M instability for a long ($k_p \sigma_z \gg k_{\beta} z$) bunch at position $\xi=z-ct$ along the bunch with peak density $n_b$, after a propagation distance $z$ in the plasma, is given by $e^\mathrm{G}$, where~\cite{bib:pukhov_prl_2011}:

\begin{equation}
\label{eq:efolding}
\mathrm{G} \cong \frac{3\sqrt{3}}{4} \left(\frac{n_b}{n_0} \left(k_p \xi\right) \left(k_{\beta} z\right)^2\right)^{1/3}.
\end{equation}
The expression for $\mathrm{G}$ from Ref.~\cite{bib:pukhov_prl_2011} is strictly valid for a flat top cylindrically symmetric narrow bunch ($k_p \sigma_r\ll 1$) with matched emittance. For our parameters ($k_p \sigma_r \simeq 0.9$) there are almost no differences between the growth rates from Ref.~\cite{bib:pukhov_prl_2011} and Ref.~\cite{bib:schroeder_prl_2011}, which is valid also for broad bunches with $k_p \sigma_r\gtrsim 1$.
%

%
Considering $\xi=\sigma_{\xi}$ and the same propagation distance $z$, Eq.~(\ref{eq:efolding}) evaluated with the parameters of Table~\ref{tab:parameters} shows that the growth rate in the lepton case is approximately 20 times larger than in the proton bunch case. The instability is therefore expected to develop over much shorter distances in the lepton case. Numerical simulations 
support this evaluation, showing cm-scale evolution distances with leptons (see below) rather than meter-scale distances with protons \cite{bib:caldwell11}.
A S-M experiment could therefore be performed 
using the long lepton bunches of early single bunch PWFA experiments \cite{bib:muggli_prl_2004} ($\propto500~\mu\mathrm{m}$) and high-density plasmas currently available for two ultra-short bunches PWFA experiments\cite{bib:hogan_njp_2010} ($\propto10^{17}~\mathrm{cm^{-3}}$). With these parameters many of the S-M physics could be tested very soon over centimeter to meter-scale plasma lengths and with most of the diagnostics also available (see below). In addition, seeding of the S-M instability could also be tested. Collimation techniques \cite{bib:muggli_prl_2008} that will be used to shape the short electron bunch into a drive/witness train \cite{bib:hogan_njp_2010} can be used to create a sharp rising edge, shorter than the plasma wavelength, in the long bunch current distribution. A bunch with a sharp edge excites a low amplitude wake in a controlled manner making S-M more predictable. This is crucial to deterministically inject a witness bunch into the accelerating phase of the wakefields. 

Since short ($\approx20~\mu$m) electron and positron bunches will also be available, acceleration of a witness bunch could be tested. Note that generation of wakefields and thus seeding of the instability could be accomplished by a short and intense laser pulse. Seeding by injecting the short electron or positron bunch in front of the long (opposite charge) bunch could be explored. Moreover, with the pre-ionized plasma source that will be available \cite{marshjoshi} the sensitivity of the instability development and saturation to plasma density gradients could be determined \cite{schroeder12}. Plasma density variations larger than the inverse number of beamlets created along the bunch by the S-M instability are expected to detune the instability and decrease the wakefields amplitude. In addition, the development of the instability along the plasma could be studied by varying the plasma length.

\begin{table}[h!]
\centering
\begin{tabular}{lll}
\hline\hline
Parameter & PDPWFA & PWFA \\
\hline
$\sigma_r [\mathrm{\mu m}]$ & 200 &  10 \\
$\sigma_r [c/\omega_p]$ & 0.4 & 0.9  \\
\hline
$\sigma_{\xi} [\mathrm{cm}]$ & 10 & $5\times 10^{-2}$  \\
$\sigma_{\xi} [c/\omega_p]$ & $188$ & $45$  \\
\hline
$\gamma_0 $  & $480$ & $4\times 10^4$  \\ 
$\gamma_0 mc^2/e [\mathrm{GeV}]$ & 450 & 20.5 \\
\hline
$N_{\mathrm{part}} $  & $11\times 10^{10}$ & $2\times 10^{10}$  \\
$n_0 [\mathrm{cm}^{-3}]$ & $10^{14}-10^{15}$ & $2.3\times10^{17}$ \\
$n_b/n_0$ & $2.0\times 10^{-2}$ & $10^{-1}$ \\
\hline
$L^{\mathrm{plasma}} [\mathrm{m}] $  & 5 & 1  \\
$L^{\mathrm{plasma}} [c/\omega_p] $  & $9\times 10^3$ & $9\times 10^4$  \\
\hline
$\epsilon_N[\mathrm{mm}\cdot\mathrm{mrad}]$ & 3.83 & 50 \\ 
\hline\hline
\end{tabular}
\caption{\label{tab:parameters} Typical parameters for the long proton bunch experiments (PDPWFA)\cite{bib:pukhov_prl_2011} and for the long electron and positron bunches experiments considered here (PWFA). The particle bunch length and transverse size are given by $\sigma_{\xi}$, and $\sigma_r$, respectively. In addition, the plasma length is denoted by $L^{\mathrm{plasma}}$ and $\epsilon_N$ is the normalized emittance.}
\label{Table:parameters}
\end{table}

\section{Simulation parameters}
\label{sec:sim_param}

We have performed 2D cylindrically symmetric and 3D numerical simulations using the fully relativistic, massively parallel particle-in-cell code OSIRIS~\cite{bib:fonseca_book}. The 2D (3D) simulations use a moving window that propagates at $c$, with resolutions $k_p\Delta z = k_p \Delta r = 0.0375$ ($k_p \Delta z = k_p \Delta x = k_p \Delta y = 0.2$) for the longitudinal and transverse cell sizes, with $2\times2$ plasma and beam particles per cell. We use quadratic particle shapes, and current smoothing and compensation. Here $k_p=\omega_p/c$ is the wake wavenumber. In the simulations the bunch density profile is:

\begin{equation}
\label{eq:electron_beam}
\frac{n_b}{n_0} =\frac{1}{2} \frac{n_{b0}}{n_0} \left(1+\cos{\left[\sqrt{\frac{\pi}{2}}\frac{\left(\xi-\xi_0\right)}{\sigma_{\xi}}\right]}\right)\exp\left[-\frac{r^2}{2 \sigma_r^{2}}\right],
\end{equation}
for $0<\xi-\xi_0<2 \sigma_{\xi} \sqrt{2 \pi}$. The bunch center is at $\xi_0=\sigma_{\xi} \sqrt{2 \pi}$. The bunch and plasma physical parameters are those of the PWFA case of Table~\ref{Table:parameters}, unless otherwise specified. The incoming bunch has zero initial energy spread. We do not present simulations for the proton bunch case, and refer to previous~\cite{bib:pukhov_prl_2011} results.

\section{Competition between hosing-like and self-modulation instabilities}
\label{sec:hosing_vs_sm}

\begin{figure}
\includegraphics[width=\columnwidth]{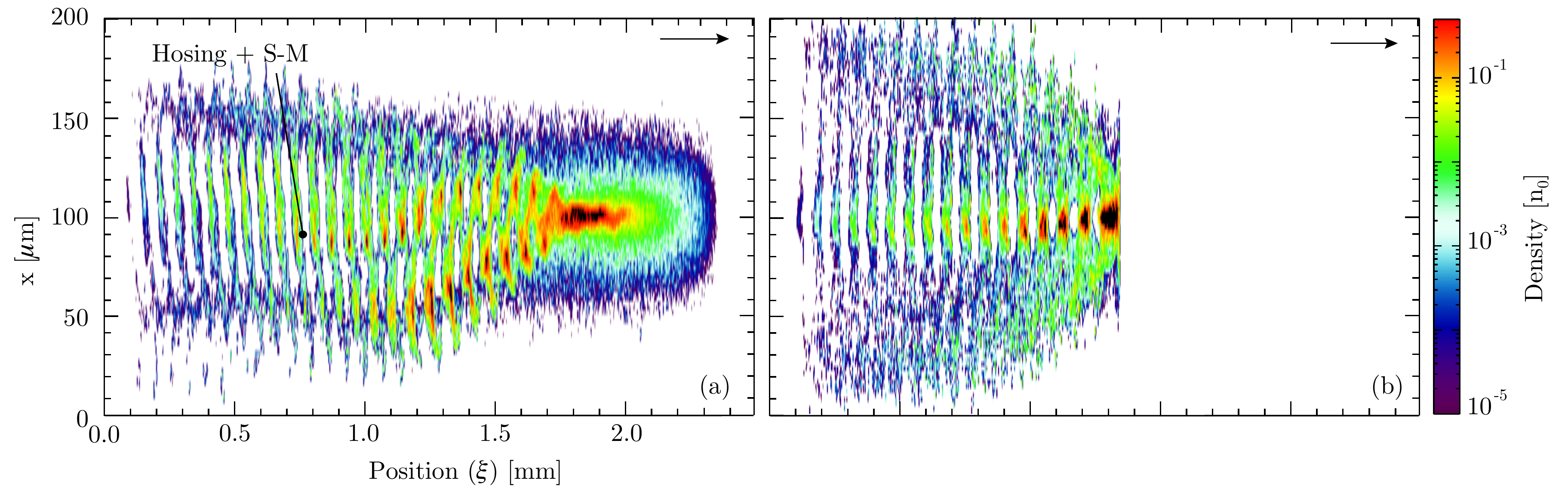}
\caption{\label{fig:3d} Results of 3D OSIRIS simulations illustrating the logarithm of the bunch density after propagation in $7.5~\mathrm{cm}$ of plasma. Figure (a) shows the competition between the hosing instability and S-M instability leading to breakup in the full length bunch case. Figure (b) shows that the hosing instability can be effectively mitigated by using a hard-cut bunch (cut in the middle of the bunch in this case) that seeds the S-M instability, enabling it to grow, saturate and further drive large amplitude wakefields over long plasma lengths. The bunches propagate towards the right as indicated by the arrows. Note that in these simulations $n_b/n_0=5\times10^{-2}$ since $N_{\mathrm{part}}=1\times10^{10}$.}
\end{figure}

The 3D simulations show that the early propagation of the full length SLAC bunch is dominated by the competition between the S-M and the hosing-like instability. The seed for the hosing in the simulations is the slight radial asymmetries in the transverse distribution resulting from the initial macro-particle distribution. Actual particle bunches may have incoming correlations (in space and momentum) and tilts that could seed the instability at a higher level. It is likely that these will be the dominant seed for hosing-like instabilities in experiments.
The hosing-like instability leads to bunch break up shortly after the beginning of the plasma (Fig.~\ref{fig:3d}a), i.e., over a length scale comparable to that of the saturation of the S-M instability ($\approx5~\mathrm{cm}$, see Section\ref{sec:wake}). However, these simulations show that hosing can be mitigated by using bunches with a short rise time in comparison to the plasma wavelength $\lambda_p = 2 \pi c/\omega_p$ (Fig.~\ref{fig:3d}b). This occurs because the short rise time of the beams provides an initial wakefield that can effectively seed the S-M instability so that it grows substantially before hosing occurs. Seeding mechanisms are important to stabilize the accelerating structures of the self-modulated PWFA~\cite{bib:kumar_prl_2010} providing a controllable noise source and relative phase for the growth of the instability.

\section{Electron and positron bunch self-modulation}
\label{sec:ep}

Now that we have established that seeding of the S-M instability can mitigate the growth of the hosing instability, we use 2D cylindrically symmetric OSIRIS simulations with assymetric bunches with rapid rise times to investigate the evolution of the wakefields over meter-long plasmas. We note that in the case of a bunch  with a sharp rise time the initial wakefield amplitude is much higher than that of the full bunch. However, the number of particles available to drive the instability is lower and the effective bunch length shorter. Therefore, the evolution of the self-modulated wakefields results from a balance between the initial amplitude of the seeding wakefields, and the corresponding number of bunch particles available to drive the wakefield. This balance may be investigated in detail in experiments with long electron and positron bunches and is briefly addressed in Section~\ref{sec:halfthreequartercut} for the electron bunch case. 

\begin{figure}
\begin{center}
\includegraphics[width=\columnwidth]{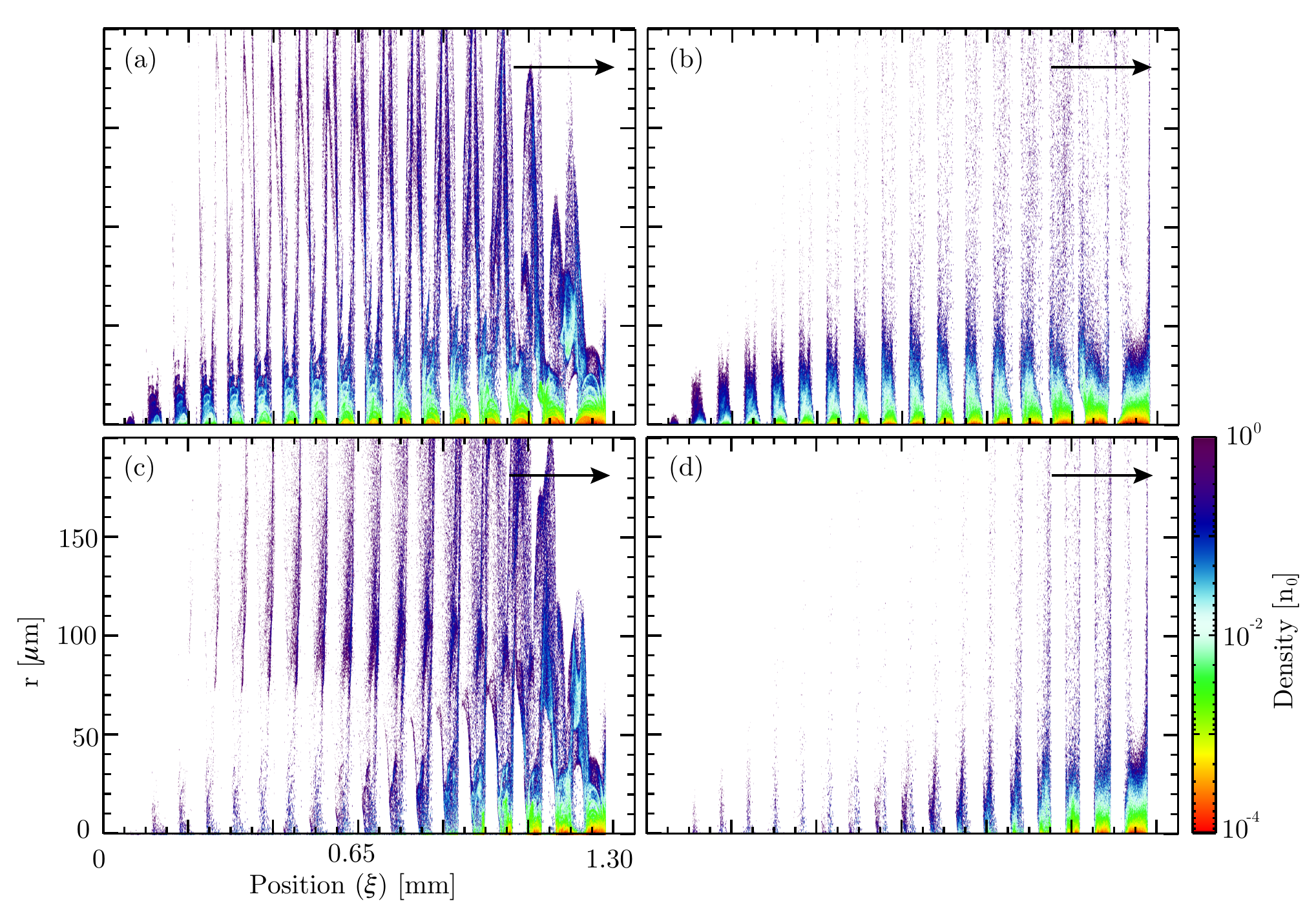}
\caption{\label{fig:selfmodulation} Cylindrically symmetric OSIRIS simulation results showing the electron bunch density after z=10 cm (a) and z = 1 m (b) and the positron bunch density after z=10 cm (c) and z = 1 m (d). The bunches propagate towards the right as indicated by the arrows. In both cases there is a hard-cut in the middle of the bunch at $\xi = 1.27~\mathrm{mm}$. Note that the density is shown in logarithmic scale.
}  
\end{center}
\end{figure}

We start with the case of a half-cut bunch where the density profile is given by Eq.~(\ref{eq:electron_beam}) with $0<\xi<\xi_0$. Figure~\ref{fig:selfmodulation} shows the electron bunch density after a propagation distance of (a) $z=10~\mathrm{cm}$ and (b) $z=1~\mathrm{m}$ (length comparable to that of the available plasma). These show that the S-M leads to the formation of several beamlets ($\propto17$ in this case) that can be clearly identified in Fig.~\ref{fig:selfmodulation}~a. Electrons initially propagating in (de)focusing regions of the wakefield are (de)focused, thus enhancing the initial (de)focusing plasma fields providing the feedback for the transverse two-stream instability~\cite{bib:kumar_prl_2010}. However, just as for laser pulse self-modulation only half of the beamlet resides completely in a focusing region in the linear phase~\cite{bib:mori_ieee_1997}.

The positron bunch densities at $z=10~\mathrm{cm}$, and at $z=1~\mathrm{m}$, are shown in Figs.~\ref{fig:selfmodulation} c and d. Figure~\ref{fig:selfmodulation} shows that the S-M instability associated with positron and electron bunches can differ due to nonlinear effects. The number of bunches is similar to that in the electron bunch case, however, there are already much less particles in the back of the bunch at $z=10~\mathrm{cm}$. Note that at this distance in the plasma the non-linear regime is already reached (see next Section). After the propagation of $z=1~\mathrm{m}$, the back of the positron bunch is strongly defocused, and the total number of positrons available to drive the wakefields is much lower than in the electron bunch case. Simulations show that after $1~\mathrm{m}$ in the plasma $20\%$ of the initial number of positrons and $50\%$ of the number of electrons are still driving the wake. Defocused particles typically leave the plasma wave at a relatively small angle, lower than 5 mrad.

\section{Wakefield excitation}
\label{sec:wake}

\subsection{Longitudinal wakefields}

\begin{figure}
\includegraphics[scale=0.8]{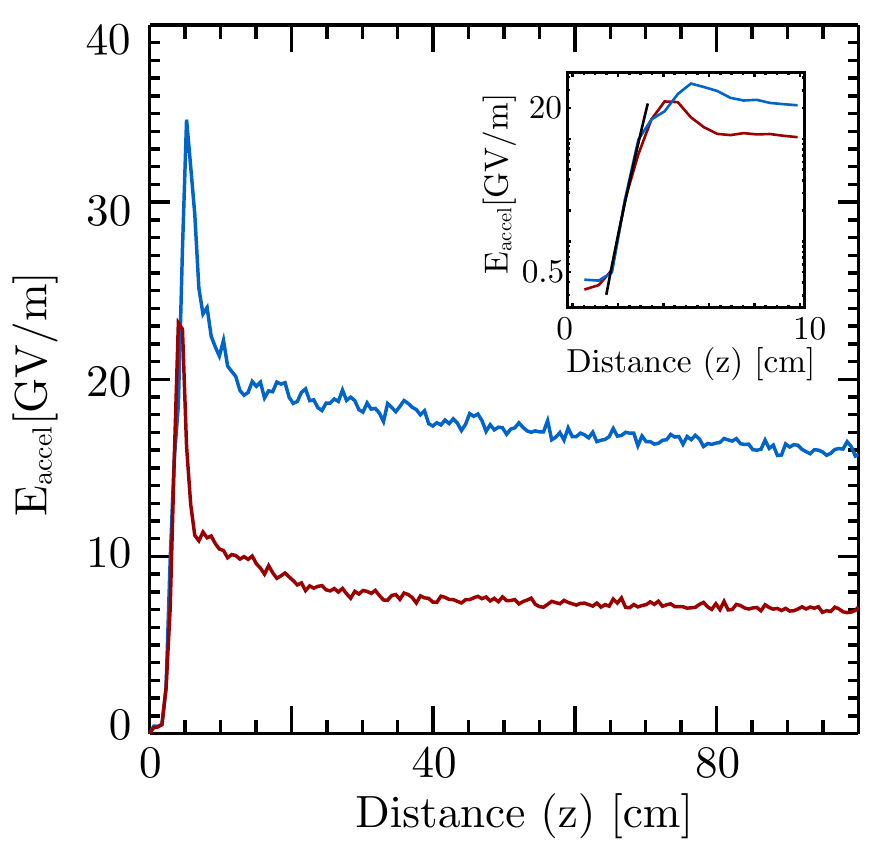}
\caption{\label{fig:ez}
Maximum accelerating field $E_{\mathrm{accel}}$ at $r=\sigma_r(z=0)/2$ (2D simulations) driven by the electron (blue line) and positron (red line) bunch along the plasma. The inset shows the logarithm of the fields along the first few centimeters, where the electric field grows exponentially in the linear regime of the instability and then saturates. The black line in the inset is the theoretical result for the growth rate (Eq.~(\ref{eq:efolding})).} 
\end{figure}

Figure~\ref{fig:ez} illustrates the amplitude of the peak accelerating fields as a function of the propagation distance in the plasma. It shows that the initial stage of the instability is similar for electron and positron bunches, in agreement with theoretical results \cite{bib:kumar_prl_2010,bib:pukhov_prl_2011,bib:schroeder_prl_2011} since the instability is in its linear growth phase. However, they differ once the wakefield amplitude reaches $\approx20~$GV/m ($z\simeq5~\mathrm{cm}$), corresponding to the excitation of non-linear wakefields close to the blowout regime. For $z\gtrsim 5~\mathrm{cm}$ the accelerating gradient driven by the positron bunch decreases to nearly one-half of that of the electron bunch. Comparison between Eq.~(\ref{eq:efolding}) 
($\mathrm{G}$ at $\xi=\sigma_z$) and the simulation results is shown in the inset of Fig.~(\ref{fig:ez}), and we find very good agreement between theory and simulations for $z<4~\mathrm{cm}$, i.e. in the linear regime.

\subsection{Transverse wakefields}

\begin{figure}
\includegraphics[width=\columnwidth]{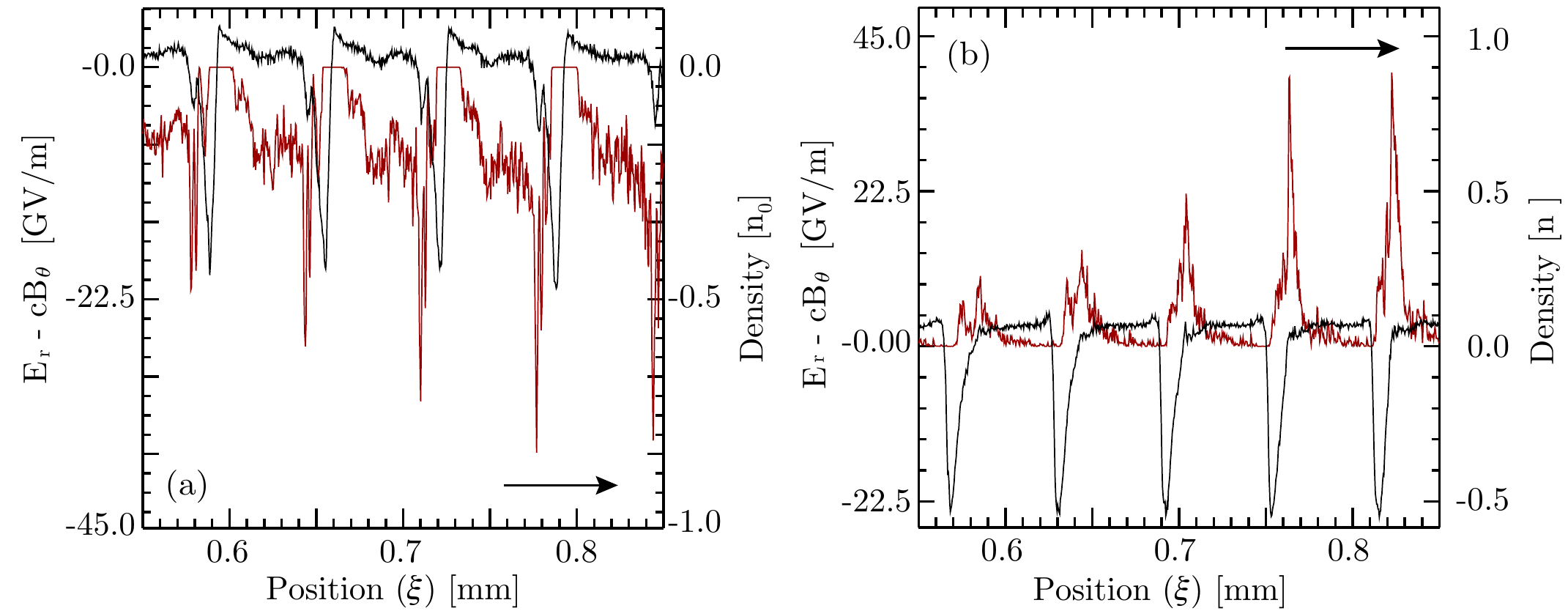}
\caption{\label{fig:er} Radial focusing ($E_r-c B_{\theta}$) field (black line, evaluated at $r=\sigma_r(z=0)/2$, 2D simulation) and bunch charge density (red line) at $z=2.5~\mathrm{cm}$ for (a) the electron and (b) the positron bunch. Only the $\xi=0.55~\mathrm{mm}$ to $\xi=0.85~\mathrm{mm}$ fractions of the bunches are plotted. The bunch propagates towards the right, as indicated by the arrows.
}  
\end{figure}

The reason for the evolution of the accelerating field can be further explored by investigating the transverse radial electric field. Figure~\ref{fig:er} shows the lineout of the radial plasma focusing force $E_r-c B_{\theta}$ (evaluated at $r=\sigma_{r0}/2$)
where $\sigma_{r0}$ is the initial spot-size, superimposed with the lineout of the self-modulated electron (Fig.~\ref{fig:er}a) and positron  (Fig.~\ref{fig:er}b) bunch densities. These fields do not vary sinusoidally along $z$, as expected in the linear regime, even after such a short propagation distance, confirming that non-linear, large amplitude wakefields are excited. 
The nonlinear regime of the plasma wakefield is quickly reached in the seeded case considered here because the initial wakefield amplitude (the seed) already reaches a few percents of the cold plasma 
wavebreaking 
field amplitude. In both cases, the leading edge of each nonlinearly self-modulated beamlet propagates in regions of defocusing fields ($E_r-c B_{\mathrm{\theta}}<0$ for electrons, $E_r-c B_{\mathrm{\theta}}>0$ for positrons), leading to the gradual erosion of the head of each beamlet (cf. Fig.~\ref{fig:selfmodulation}b). As a result the accelerating wakefield gradually decreases as it propagates (increasing z), which is consistent with Fig.~\ref{fig:ez}. In the positron bunch scenario (Fig.~\ref{fig:er}b), as non-linear wakes are excited, the fields that are focusing for positrons ($E_r-c B_{\mathrm{\theta}}<0$ for $r>0$) are limited to the regions where the plasma electrons cross the axis in the back of each plasma period. As a result the beamlets are shorter in the positron case and more positrons are defocused. Hence the accelerating field is lower than in the electron bunch case.

\subsection{Energy gain and loss}

\begin{figure}
\includegraphics[scale=0.8]{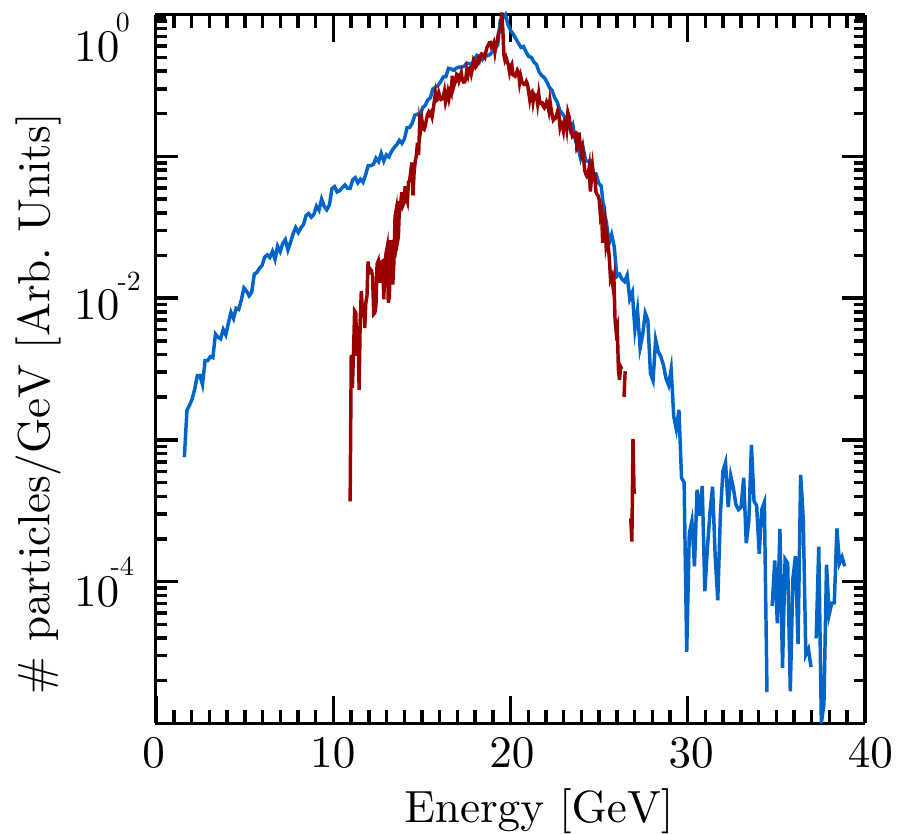}
\caption{\label{fig:energyspectra} Electron (blue line) and positron (red line) bunch energy spectra after propagation of $z=1~\mathrm{m}$ in the plasma. Both spectra are normalized to their peak value.}  
\end{figure}

The electron bunch energy spectrum at $z=1~$m (Fig.~\ref{fig:energyspectra}) shows that the initial energy varies by up to $15~\mathrm{GeV}$ (loss at the $1\%/\mathrm{GeV}$ level of the peak value). For positrons, $8~\mathrm{GeV}$ energy loss at the 1\%/GeV level occurs. In both cases energy gain on the order of $5~\mathrm{GeV}$ is observed. For the electron bunch case some beam particles gain up to $\sim 20~\mathrm{GeV}$. The broad continuous energy spectra are the result of the bunch particles sampling all the phases of the wakefield. 

It is interesting to note that as the blowout regime is reached, the peak wakefield amplitudes driven by the long, self-modulated bunches are comparable to that obtained in recent PWFA experiments using compressed bunches of length approximately equal to that of the beamlets, but containing the full charge ($\approx50~\mathrm{GV/m}$)~\cite{bib:hogan_prl_2005,bib:blumenfeld_nat_2007,bib:muggli_njp_2010}. This shows the effectiveness of the self-modulated bunch in driving large amplitude accelerating fields through the resonant excitation of plasma wakefields. Uncompressed lepton bunches are more easily produced in radio-frequency linacs and using a seeded self-modulated PWFA (rapid rise) may reduce the complexity of the drive bunch linac. However, since only about half the particles participate in the wake excitation and the wakefield amplitude varies along the bunch the energy transfer efficiency will be proportionally smaller.

\section{Evolution of the wakefields along the plasma}
\label{sec:wakefieldevolution}

\subsection{Phase velocity}

With available parameters (See Table~\ref{Table:parameters}), the PDPWFA would operate in the linear regime~\cite{bib:caldwell_ppcf_2011} of the PWFA. Unlike the PDPWFA, the self-modulated PWFA can lead to the generation of non-linear wakes close to the blowout regime~\cite{bib:lu_2006}. Because in the blowout regime the non-linear plasma wavelength differs from that of the linear regime, the phase velocity of the self-modulated wakefields suffers from additional spatial and temporal variations. However, these issues can be overcome by splitting the S-M and the acceleration plasma sections. In an accelerator scheme, the self-modulation of the drive bunch may occur in a short S-M plasma section, where the wake phase velocity is slow when compared to that of the drive bunch \cite{bib:pukhov_prl_2011,bib:schroeder_prl_2011} (see Section~\ref{sec:wakefieldevolution}). The acceleration would then occur in a following plasma section, where the wake phase velocity is close to that of the drive bunch. This scheme would allow for the external injection and trapping of a low energy witness bunch followed by its acceleration in the relativistic wakefields. Note that simulations show that external injection at an angle with respect to the drive bunch trajectory may lead to a narrow final energy spread even when the injected bunch is long compared to the plasma wavelength \cite{bib:pukhov_prl_2011}.

\begin{figure}
\includegraphics[width=\columnwidth]{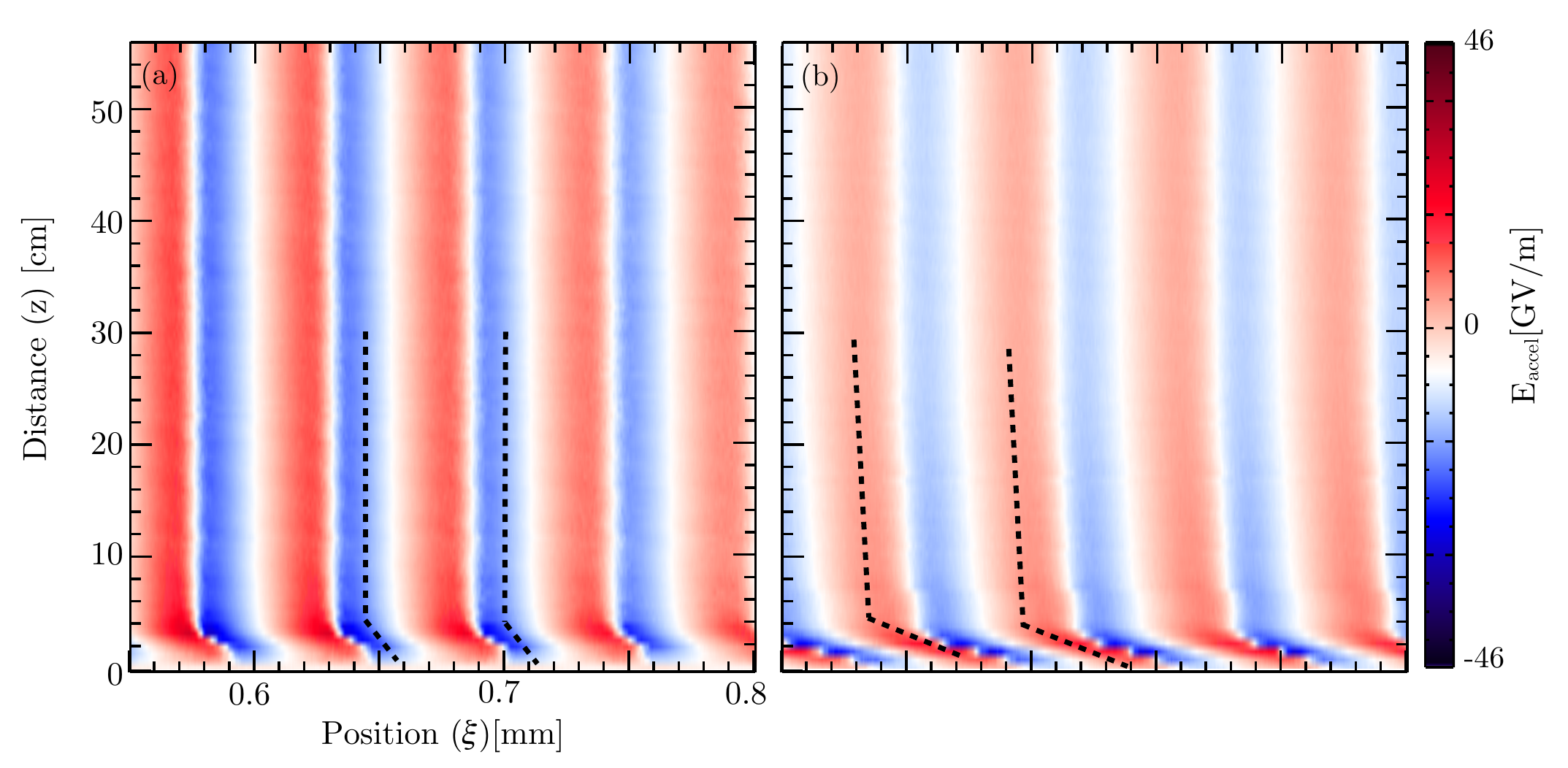}
\caption{\label{fig:vphi} Evolution of the accelerating wake structure ($E_{\mathrm{accel}}$ field) within the bunch between $\xi = 0.55~\mathrm{mm}$ and $\xi = 0.8~\mathrm{mm}$ as a function of the propagation distance for the electron bunch case, (a) and for the positron bunch case (b). The dashed lines show the approximate \emph{trajectories} of the peak accelerating field.}  
\end{figure}

The qualitative evolution of the wakefield phase velocity of lepton bunches can be examined through Fig.~\ref{fig:vphi}. It illustrates the evolution of the axial accelerating field amplitude driven by the electron (Fig.\ref{fig:vphi} a) and by the positron (Fig.\ref{fig:vphi} b) bunch. The phase velocity of the plasma waves $v_{\phi}$ is deduced from the slope of the longitudinal electric field ($E_z$) in the $z-\xi$-plane. The slope is given by $\frac{\delta z}{\delta\xi}=\frac{c}{v_{\phi}-c}$. Thus, a more (less) vertical slope corresponds to a faster (slower) phase velocity.  Two regions can be identified on Fig.~\ref{fig:vphi}. First, before the saturation of the S-M instability, for $z\lesssim 5~\mathrm{cm}$  the wake phase velocity is significantly lower than the velocity of the beam $v_b\simeq c$. This is the region corresponding to the S-M plasma section. Second, after the saturation of the S-M instability, for $z>5~\mathrm{cm}$, the wake phase velocity is close to $v_b\sim c$ and dephasing is not an issue~\cite{bib:pukhov_prl_2011}. This is the acceleration plasma section. Figure~\ref{fig:vphi} shows that in both regions $v_{\phi}$ is larger with the electron than with the positron bunch, further indicating that electron bunches are advantageous for the acceleration of an externally injected witness particle bunch. Simulations show that these differences in self-modulated wakefields phase velocity are due to the smaller focusing field regions for positrons than for electrons in the blowout regime. This leads to faster erosion of the positron beamlets head (slower phase velocity) and lower overall wakefield amplitudes.

\subsection{1/2 versus 3/4 cut-bunch seeding}
\label{sec:halfthreequartercut}

\begin{figure}
\begin{center}
\includegraphics[width=\columnwidth]{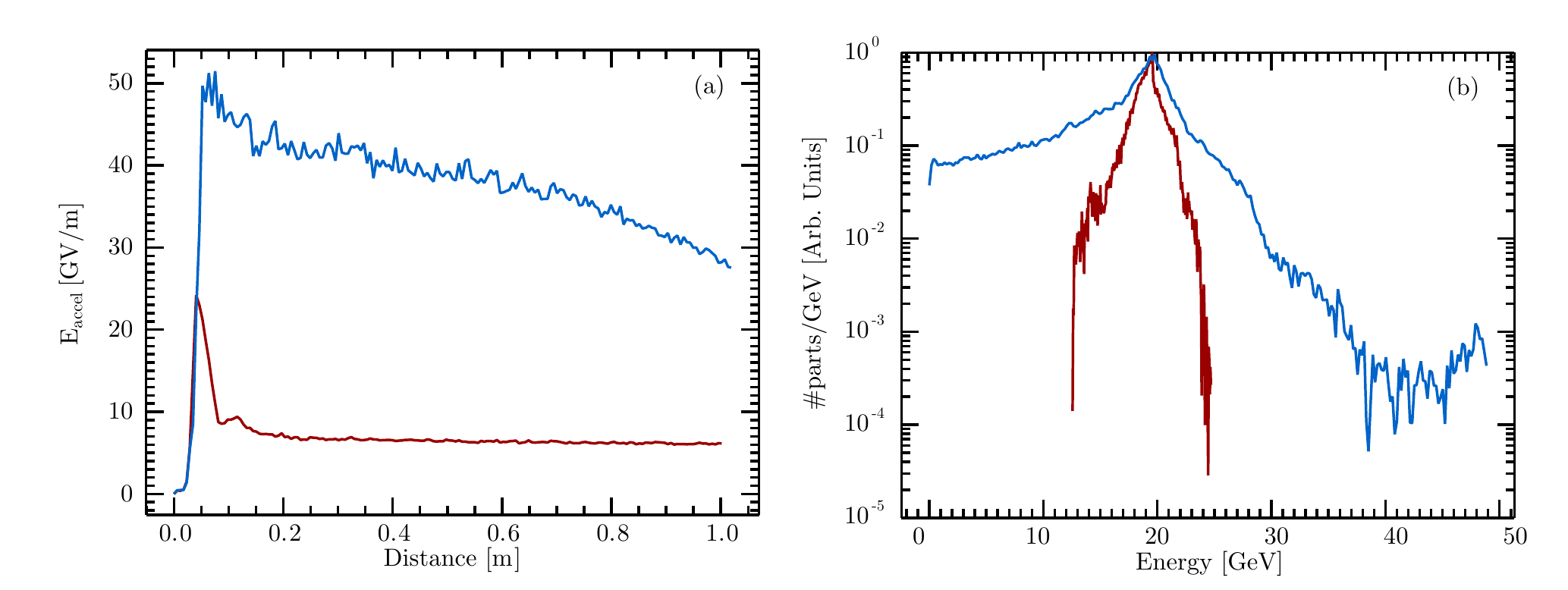}
\caption{\label{fig:cut} (a) Maximum accelerating fields for the 3/4-cut bunch using electron (blue line) and positron (red line) bunches. (b) corresponding electron (blue line), and positron (red line) bunch spectra after 1 meter propagation.}  
\end{center}
\end{figure}

We also performed simulations to show how the results change when the location of the rapid rise was earlier in the bunch. We considered a bunch with a hard cut density profile given by Eq.~(\ref{eq:electron_beam}) with $0 < \xi-\xi_0 < 3/2 \sigma_{\xi} \sqrt{2 \pi}$ (3/4-cut bunch), rather than with $0 < \xi-\xi_0 < \sigma_{\xi}/\sqrt{2 \pi}$  (half-cut bunch), since this can be tested at SLAC FACET. The simulation results shown in Fig.~\ref{fig:cut}a reveal that the 3/4-cut bunch provides larger accelerating gradients in comparison to the half-cut bunch because it carries more charge and is longer, despite the lower initial amplitude of the seeding compared to the half-cut case. As a result, the corresponding particle energy spectra of Fig.~\ref{fig:cut}b show energy variations in excess of 20 GeV at the 1\%/GeV level for electrons, and 6~GeV for positrons (compare with Figs.~\ref{fig:ez} and \ref{fig:energyspectra}). Note however that the longer 3/4-cut bunch is more sensitive to the hosing-like instability than the 1/2-cut bunch. Its propagation stability will be studied using 3D simulations in a future publication. 

\section{Observation of the S-M instability}
\label{sec:challenges}

\begin{figure}
\begin{center}
\includegraphics[scale=0.7]{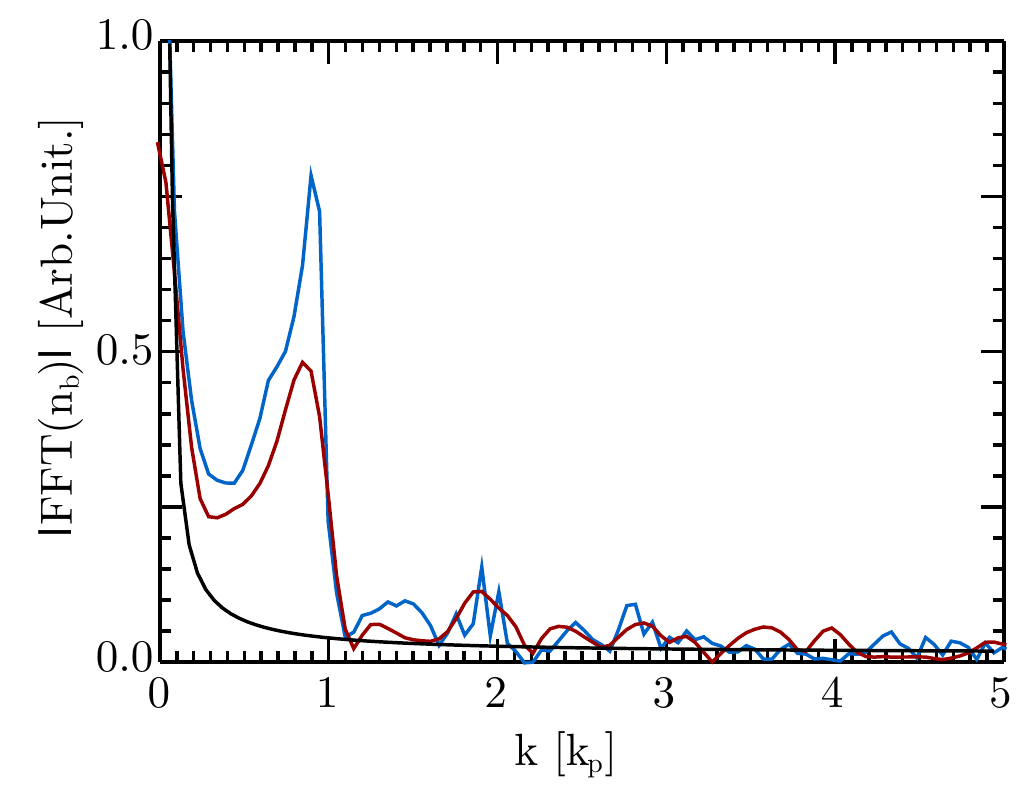}
\caption{\label{fig:FFT} Absolute value of the fast Fourier transform of the electron (blue line) and positron (red line) bunch on-axis density after $1~\mathrm{m}$ of plasma to be compared with that of the unmodulated incoming bunch (black line). The peak at $k=k_p$ corresponds to the S-M of the bunches at the plasma wavelength. The peak at the first harmonic ($k=2k_p$) is visible in the positron bunch case, for which the beamlets are shorter.}  
\end{center}
\end{figure}

Since simulations indicate that the S-M and the excitation of wakefields reach the non-linear regime with available electron and positron bunches, we briefly explore how S-M and its effects on the bunches could be diagnosed in an experiment. The S-M could be measured with coherent radiation detection techniques. With the parameters considered here the period of the S-M is in the $70~\mu$m range. Therefore, the coherent transition radiation (CTR) emitted by the self-modulated train when traversing a thin metallic foil is in the $4~\mathrm{THz}$ frequency range and below. Emission of CTR at frequencies $>1~\mathrm{THz}$ when the bunch travels in the plasma as opposed to $<1~\mathrm{THz}$ without plasma ($\sigma_z=500~\mathrm{\mu m}$) would be a clear indication that S-M occurred. Such an integrated CTR energy measurement has already been successfully used to monitor the relative length of ultra-short electron bunches on a shot-to-shot basis \cite{bib:muggli_njp_2010}. CTR auto-correlation measurements could also be used to determine the S-M period in an average sense, assuming that the development of the instability is sufficiently reproducible. Information about the bunch modulation depth and beamlets actual length could in principle be obtained from the spectrum of the coherent radiation. Fourier transforms of the electron and positron bunch longitudinal density profiles after $1~\mathrm{m}$ of plasma are shown in Fig.~\ref{fig:FFT}. These could be obtained from single shot Smith-Purcell diagnostic~\cite{bib:smith_pre_1953} and show clear peaks at the plasma wavelength ($k=k_p$ in Fig.~\ref{fig:FFT}). Peaks at the second harmonic ($k=2k_p$) are also visible in Fig.~\ref{fig:FFT}, especially in the positron bunch case, characterized by shorter beamlets (see Fig.~\ref{fig:selfmodulation} and Fig.~\ref{fig:er}). Unless the instability is seeded, single shot diagnostic are necessary to capture the instability-aspect of the self-modulation. 
Note however that in the case of ultra-relativistic bunches considered here collimation of the defocused particles to yield a real bunch current modulation at the diagnostic will be necessary. This is due to the fact that with high-energy particle the coherent radiation is essentially emitted in the forward direction, significantly decreasing the dependency of the emitted energy or of the emitted spectrum on the transverse size modulation.

In addition, energy changes of the bunch particles could be easily measured since the values for energy gain and loss obtained from simulations are well within the energy measurement limits of previous single, short bunch PWFA experiments \cite{bib:hogan_prl_2005,bib:blumenfeld_nat_2007,bib:muggli_njp_2010}. Finally the focusing/defocusing effect of the S-M instability on the bunch particles could also be detected by observing the bunch transverse size, a short distance downstream from the plasma exit, using for example optical transition radiation imaging~\cite{bib:hogan_prl_2005,bib:transverse}.

\section{Conclusions}
\label{sec:conclusions}

In conclusion, this work reveals that uncompressed electron and positron bunches can be used to observe self-modulation of ultra-relativistic particle beams for the first time in cm to meter-long plasmas. With these bunch parameters one can access and understand the physics of the self-modulated plasma wakefield accelerator, in the linear and in the blowout regime.
Different seeding mechanisms for the S-M instability can be tested. The hosing-like instability of self-modulated beams can also be tested by using full beams, and beams with rapid rise times that are predicted to mitigate its appearance. This is also confirmed by numerical simulations. Other mechanisms that can lead to the suppression of the S-M instability, including the motion of the background plasma ions~\cite{bib:vieira_tosubmit}, could also be investigated. The possible experiments present outstanding opportunities to unravel the physical mechanisms associated with the propagation of long particle beams in dense plasmas.

\acknowledgements

Work partially supported by FCT (Portugal) through grants SFRH/BPD/71166/2010, PTDC/FIS/111720/2009, and CERN/FP/116388/2010; European Research Council (ERC-2010-AdG Grant 267841 ACCELERATES); UC Lab Fees Research Award No. 09-LR-05-118764-DOUW, the US DoE under DE-FC02-07ER41500, DE-FG02-92ER40727 and DE-FG02-92ER40745, and the NSF under NSF PHY-0904039 and PHY-0936266. Simulations were performed on the IST Cluster at IST, on the Jugene supercomputer under a ECFP7 and a DEISA Award, and on Jaguar computer under an INCITE Award.


\begin{thebibliography}{40}
\bibitem{bib:tajima_prl_1979} T. Tajima and J. M. Dawson, Phys. Rev. Lett. \textbf{43}, 267 (1979).
\bibitem{bib:chen_prl_1985} P. Chen, J. M. Dawson, R. W. Huff, and T. Katsouleas, Phys. Rev. Lett. \textbf{54}, 693 (1985). 
\bibitem{bib:patel_nature_2007}N. Patel, Nature \textbf{449}, 133 (2007).
\bibitem{bib:caldwell_natphys_2009} A. Caldwell, K. Lotov, A. Pukhov, and F. Simon, Nat. Phys. 5, 363 (2009).
\bibitem{bib:kumar_prl_2010} N. Kumar, A. Pukhov, and K. Lotov, Phys. Rev. Lett. \textbf{104} 255003 (2010).
\bibitem{bib:esarey_prl_1996} E. Esarey, J. Krall, P. Sprangle, Phys. Rev. Lett. \textbf{72} 2887 (1994).
\bibitem{bib:mori_ieee_1997} W.B. Mori, IEEE Trans. Plasma Sci. \textbf{33} 1942 (1997).
\bibitem{bib:fil} R. Keinigs, M. E. Jones, Phys. Fluids \textbf{30}, 252 (1987).
\bibitem{bib:hogan_njp_2010} M. J. Hogan, T. O. Raubenheimer, A. Seryi, P. Muggli, T. Katsouleas, C. Huang, W. Lu, W. An, K. A. Marsh, W. B. Mori, C. E. Clayton, and C. Joshi, New J. Phys. \textbf{12}, 055030 (2010).
\bibitem{bib:whittum_prl_1991} D. Whittum, W. M. Sharp, S. S. Yu, M. Lampe, and G. Joyce, Phys. Rev. Lett. \textbf{67}, 991 (1991).

\bibitem{bib:duda_prl_1999} B. J. Duda,  R. G. Hemker, K. C. Tzeng, and W. B. Mori, Phys. Rev. Lett. \textbf{83}, 1978 (1999).
\bibitem{bib:duda_pre_2000} B. J. Duda and W.B. Mori, Phys. Rev. E. \textbf{61}, 1925 (2000). 
\bibitem{bib:lu_2006} J. Rosenzweig, B. Breizman, T. Katsouleas, and J. J. Su, Phys. Rev. A \textbf{44}, R6189-6192 (1991); W. Lu, C. Huang, M. Zhou, W. B. Mori, and T. Katsouleas, Phys. Rev. Lett. \textbf{96}, 165002 (2006); W. Lu, M. Tzoufras, C. Joshi, F. S. Tsung, W. B. Mori, J. Vieira, R. A. Fonseca, and L. O. Silva, Phys. Rev. ST Accel. Beams \textbf{10}, 061301 (2007).

\bibitem{bib:lu_pop_2005} W. Lu, C. Huang, M. M. Zhou, W. B. Mori, and T. Katsouleas, Phys. Plasmas \textbf{12}, 063101 (2005).


\bibitem{bib:pukhov_prl_2011} A. Pukhov,  N. Kumar, T. TŸckmantel, A. Upadhyay, K. Lotov, P. Muggli, V. Khudik, C. Siemon, and G. Shvets, Phys. Rev. Lett. \textbf{107} 145003 (2011).
\bibitem{bib:schroeder_prl_2011} C. B. Schroeder, C. Benedetti1, E. Esarey, F. J. GrŸner, and W. P. Leemans, Phys. Rev. Lett. \textbf{107} 145002 (2011).
%
\bibitem{bib:caldwell11} A. Caldwell and K. V. Lotov, Phys. Plasmas \textbf{18}, 103101 (2011)
\bibitem{bib:muggli_prl_2004} B.E. Blue, C. E. Clayton, C. L. OÕConnell, F.-J. Decker, M. J. Hogan, C. Huang, R. Iverson, C. Joshi, T. C. Katsouleas, W. Lu, K. A. Marsh, W. B. Mori, P. Muggli, R. Siemann, and D. Walz, Phys. Rev. Lett. \textbf{90}, 214801 (2003); P. Muggli, B. E. Blue, C. E. Clayton, S. Deng, F.-J. Decker, M. J. Hogan, C. Huang, R. Iverson, C. Joshi, T. C. Katsouleas, S. Lee, W. Lu, K. A. Marsh, W. B. Mori, C. L. O'Connell, P. Raimondi, R. Siemann, and D. Walz, Phys. Rev. Lett.  \textbf{93}, 014802 (2004).
\bibitem{bib:muggli_prl_2008} P. Muggli, V. Yakimenko, M. Babzien, E. Kallos, and K. P. Kusche, Phys. Rev. Lett. \textbf{101}, 054801 (2008).
\bibitem{marshjoshi} K.A. Marsh, C. Joshi, private communication.
%
\bibitem{schroeder12} C. B. Schroeder, C. Benedetti, E. Esarey, F. J. GrŸner, and W. P. Leemans, Phys. Plasmas \textbf{19}, 010703 (2012). 
\bibitem{bib:fonseca_book} R. A. Fonseca, L. O. Silva, F. S. Tsung, V. K. Decyk, W. Lu, C. Ren, W. B. Mori, S. Deng, S. Lee and T. Katsouleas, Lect. Notes Comp. Sci. vol. 2331/2002, (Springer Berlin / Heidelberg), (2002).
\bibitem{bib:hogan_prl_2005} M. J. Hogan, C. D. Barnes, C. E. Clayton, F. J. Decker, S. Deng, P. Emma, C. Huang, R. H. Iverson, D. K. Johnson, C. Joshi, T. Katsouleas, P. Krejcik, W. Lu, K. A. Marsh, W. B. Mori, P. Muggli, C. L. OÕConnell, E. Oz, R. H. Siemann, and D. Walz, Phys. Rev. Lett. \textbf{95}, 054802 (2005).
\bibitem{bib:blumenfeld_nat_2007} I. Blumenfeld, C. E. Clayton, F.-J. Decker, M. J. Hogan, C. Huang, R. Ischebeck, R. Iverson, C. Joshi, T. Katsouleas, N. Kirby, W. Lu, K. A. Marsh, W. B. Mori, P. Muggli, E. Oz, R. H. Siemann, D. Walz, M. Zhou, Nature \textbf{445}, 741 (2007).
\bibitem{bib:muggli_njp_2010} P. Muggli, I. Blumenfeld, C. E. Clayton, F. J. Decker, M. J. Hogan, C. Huang, R. Ischebeck, R. H. Iverson, C. Joshi, T. Katsouleas, N. Kirby, W. Lu, K. A. Marsh, W. B. Mori, E. Oz, R. H. Siemann, D. R. Walz, and M. Zhou, New J. Phys. \textbf{12}, 045022 (2010).
\bibitem{bib:caldwell_ppcf_2011} A. Caldwell, K. Lotov, A. Pukhov and G. Xia, Plasma Phys. Control. Fusion \textbf{53}, 014003 (2011).
\bibitem{bib:smith_pre_1953} S. J. Smith and E. M. Purcell, Phys. Rev. \textbf{92}, 1069 (1953).
\bibitem{bib:transverse} C. E Clayton, B. E. Blue, E. S. Dodd, C. Joshi, K. A. Marsh, W. B. Mori, S. Wang, P. Catravas, S. Chattopadhyay, E. Esarey, W. P. Leemans, F. J. Decker, M. J. Hogan, R. Iverson, P. Raimondi, R. H. Siemann, D. Walz, T. Katsouleas, S. Lee, and P. Muggli, Phys. Rev. Lett. \textbf{88}, 154801 (2002);
\bibitem{bib:vieira_tosubmit} J. Vieira \emph{et al.} to be submitted (2012).
\end{thebibliography}
\end{document}